\documentclass[prb,preprint]{revtex4-2} 

\usepackage[utf8]{inputenc}
\usepackage[T1]{fontenc}
\usepackage{float}
\usepackage{units}
\usepackage{amsmath}
\usepackage{amsfonts} 
\usepackage{graphicx}
\usepackage{babel}
\usepackage{color}
\usepackage{comment}
\usepackage{siunitx} 
\usepackage{hyperref}

\begin{document}

\title{Coupled oscillations of the Wilberforce pendulum unveiled by smartphones}

\author{Thomas Gallot}
\email{tgallot@fisica.edu.uy}
\affiliation{Instituto de Física, Facultad de Ciencias, UdelaR,\\ Iguá 4225, 11200, Uruguay}

\author{Daniel Gau}
\email{dgau@fing.edu.uy}
\affiliation{Instituto de Física, Facultad de Ingeniería, UdelaR,\\ Av. Julio Herrera y Reissig 565, 11300, Uruguay} 

\author{Rodrigo García-Tejera}
\email{rodrigo.garcia@ed.ac.uk}
\affiliation{Centre for Regenerative Medicine, University of Edinburgh, 5 Little France Dr, Edinburgh EH16 4UU, U.K.}

\begin{abstract}
   The Wilberforce pendulum is a great experience for illustrating important properties of coupled oscillatory systems, such as normal modes and beat phenomena, in physics courses. A helical spring attached to a  mass comprises this simple but colorful device that features longitudinal and torsional oscillations, catching the eye of students and teachers, and offering a fantastic environment to study coupled oscillations at qualitative and quantitative levels. However, the experimental setups that are commonly used to simultaneously acquire both oscillation types can be cumbersome and/or costly, often requiring two synchronised cameras, or motion sensors and photo-gates integrated by an expensive analog-to-digital converter. Here, we show that a smartphone can be the perfect device for recording the oscillations of a Wilberforce pendulum composed of just an educational helical spring and a can. Our setup employs a smartphone's accelerator and gyroscope to acquire data from the longitudinal and torsional oscillations, respectively. This setup can successfully record both normal modes and beats. Furthermore, we show that our experimentally obtained time-series and power spectra are in great agreement with simulations of the motion equations. In addition, we critically analyse how to set the initial conditions to observe both normal modes and beats. We believe our study contributes to bring the striking physics of the Wilberforce pendulum closer to undergraduate classrooms with low-cost, highly-accessible materials.
\end{abstract}

\maketitle

\section{Introduction}
The oscillations of the Wilberforce pendulum have amazed people since they were first reported, at the end of the XIX century, by the British demonstrator in Physics L.R. Wilberforce \cite{wilberforce1894xliv}. In this simple device, consisting of a spiral spring attached to a mass, the linear and torsional modes of the system are coupled. When the frequencies of the two oscillation types are similar enough, the two modes exchange energy producing a fascinating beat phenomenon.  

The first reported analysis of the motion of the Wilberforce pendulum in real time was performed using an ultrasound distance sensor to detect the vertical motion of the mass \cite{debowska1999computer}. There, the authors study the beat phenomenon and analysed the motion of the system for each of the normal modes separately by applying different initial conditions to the system. This approach was then extended to include a digital video camera with the aim of detecting both the translational and rotational motion simultaneously \cite{greczylo2002using}. Torzo et al. introduced a setup that was able to detect both motions simultaneously, using an ultrasonic RTL sensor to detect the translational motion of the mass and a Non-Contact Rotation Sensor based on polarized light to measure the rotation angle \cite{torzo2004wilberforce}. More recently the detection of both movements using two cameras was also performed \cite{devaux2019cross}.  

While the original purpose of the device was the measurement of spring constants, the Wilberforce pendulum is currently used in physics teaching at Universities to illustrate a wide variety of concepts regarding oscillatory phenomena, including coupling, normal modes or beats, among others \cite{greczylo2002using,devaux2019cross,mewes2014slinky,plavcic2009resonance,pereyra2018fourier,osorio2018measuring}. In recent years the study of these fundamental concepts has been approached with the use of smartphones \cite{suarez2020normal,monteiro2019physics,monteiro2018bottle}. Smartphones are having a deep impact in physics teaching, mainly thanks to the versatility and high accessibility of these small portable laboratories. Moreover, the use of this technology has proven to have a positive impact in the interest and curiosity of students \cite{hochberg2018using}. 

In this work, we present a new approach to study the Wilberforce pendulum using simultaneously the accelerometer and gyroscope of a smartphone, which also makes a significant contribution to the pendulum's mass and inertia. This device allows for the measurement of the accelerations and angular velocities, thus shedding light into the physical processes underlying the complex motion of the mass. In section II we revisit the theory, discussing the mechanical basis of the coupling between the translational and torsional modes. We analyse the motion equations and derive analytically optimal initial conditions to observe the normal modes and perfect beatings in the acceleration signals. For the latter case, we prove that optimal beating in the accelerations is obtained through a zero-torque initial condition, rather than with the classical no-initial-rotation condition needed for beatings in the position signals. In section III we describe our experimental setup, showing how to set the Wilberforce pendulum to optimal beating conditions. We outline our numerical experiments and present experimental results in great agreement with simulations of the motion equations. The Python scripts for the experimental data analysis and numerical experiments are available on a  \href{https://github.com/RodrigoGarciaTejera/WilberforcePendulum }{GitHub repository} \cite{WFGithub}.

\section{Modelling the Wilberforce Pendulum}\label{sec:theory}

\begin{figure}[htbp]
\centering \includegraphics[width=0.7\columnwidth]{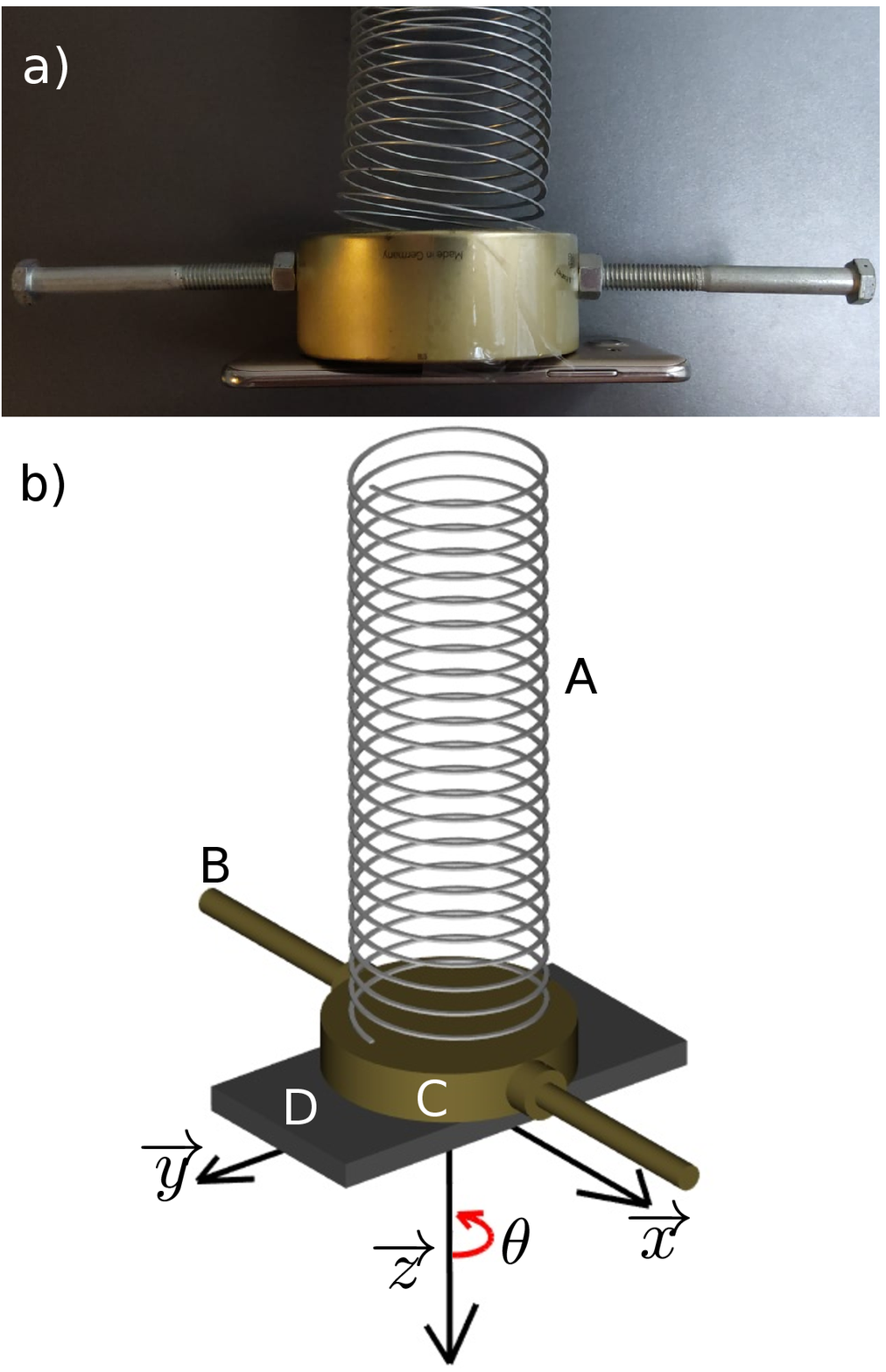}
\caption{\label{fig:Sketch} Implementation of the Wilberforce pendulum. a) Profile picture of the pendulum's mass, spring and smartphone attached. b) Schematic of the Wilberforce pendulum consisting of {\bf A.} the educational spring, {\bf B.} bolts, {\bf C.} cylindrical box, and {\bf D.} smartphone. The principal components of the motion are the translation along the $z$-axis and rotation $\theta$ around the $z$-axis.}
\end{figure}

We consider the set up depicted in Fig.~\ref{fig:Sketch}, consisting of a classical Wilberforce pendulum with a smartphone attached at the bottom. To study the motion of the system, we start by defining the natural Cartesian coordinate system ($\vec{x},\vec{y},\vec{z}$) fixed to the smartphone, as shown in Fig.~\ref{fig:Sketch}, and an angle $\theta$ representing the rotation from an initial resting point.

\subsection{Mechanical origin of the coupling}\label{MecCons}

\begin{figure}[h]
\centering \includegraphics[width=0.8\columnwidth]{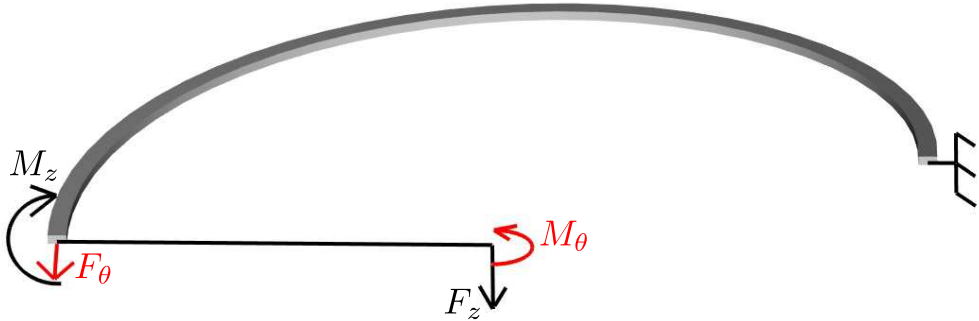}
\caption{\label{fig:Spring} Half-turn of a helicoidal spring with rectangular cross-section. The force $F_z$ along the $z$-axis associated to a translation $z$ generates a torque $M_z=2F_z/D$ along the spring's cross-section. Reciprocally, the torque $M_{\theta}$ around the $z$-axis associated to a rotation $\theta$  generates a traction-compression force $F_{\theta}=M_{\theta}D/2$ on the spring material.}
\end{figure}

The elastic constant of a spring in traction/compression is given by $k=\nicefrac{F_z}{z}$, being $F_z$ the force after a displacement $z$ along the $\vec{z}$ axis.
Similarly, the rotational elastic constant is given by $\delta=\nicefrac{M_{\theta}}{\theta}$, where $M_{\theta}$ is the torque associated to an angle variation $\theta$ around the $\vec{z}$-axis. For a helical spring with $N$ loops and diameter $D$, the spring constants can be calculated using strength mechanical theory \cite{Ancker1958,kopf1990}, arriving at
\begin{equation}
    \begin{aligned}
         k= & G\frac{I_{z}}{ D^{3}}\frac{4}{N\pi}\\
        \delta= & E\frac{I_{\theta}}{D}\frac{1}{N\pi},
    \end{aligned}
    \label{resonance}
\end{equation}
where $I_{z}$ and $I_{\theta}$ are the axial and polar moments of inertia, respectively \cite{kopf1990}. Interestingly, the traction-compression elastic constant $k$ depends on the shear modulus $G$ of the spring material \cite{Ancker1958} which characterizes a rotational strain. Reciprocally, the torsional elastic constant $\delta$ is related to the Young's modulus $E$ which characterizes a compressional strain. Both elastic moduli, and thus the elastic constants, are connected by the Poisson's ratio $\nu$: $\nicefrac{E}{G}=2\left(1+\nu\right)$. A traction-compression of the spring, (displacement along the $\vec{z}$-axis) generates a torque within the spring material and, reciprocally, a rotation of the spring generates a traction-compression (bending), see Fig.~\ref{fig:Spring}. This observation is the core essence of the Wilberforce pendulum: the compression of the pendulum creates a rotational stress and, reciprocally, the rotation of the pendulum creates a compressional stress in the spring material, thus coupling the rotational and translational motions.

Considered independently, the translational and rotational motions have resonance frequencies, $\omega_{z}$ and $\omega_{\theta}$, that depend on the elastic constants and the mass $m$ and moment of inertia $I$ about the $\vec{z}$-axis, namely
\begin{equation}
\begin{aligned}
\omega_{z}= &\sqrt{\nicefrac{k}{m}}\\
\omega_{\theta}= & \sqrt{\nicefrac{\delta}{I}}.
\end{aligned}
\label{resonances}
\end{equation}
Such resonant frequencies cannot be measured directly due to the mechanical coupling between rotation and traction-compression; only the coupled system resonances can be measured. Moreover, even if $m$ and $I$ can be estimated independently, estimating the elastic constants $k$ and $\delta$ can be challenging due to the coupling.

\subsection{Motion equations \label{subsec:motionequations}}


Following Berg and Marshall\cite{berg1991}, we consider a linear coupling between the resonators to describe the system dynamics in the absence of attenuation
\begin{equation}
\begin{cases}
m\ddot{z}+kz+\frac{\epsilon}{2}\theta=0\\
I\ddot{\theta}+\delta\theta+\frac{\epsilon}{2}z=0
\end{cases},
\label{Eqmotion}
\end{equation}
where $\epsilon$ denotes the coupling strength, and $z$ and $\theta$ the vertical and angular coordinate of the pendulum, respectively, measured from their equilibrium points. Imposing the normal mode conditions  $z\left(t\right)=A_{z}e^{i\omega t}$ and $\theta\left(t\right)=A_{\theta}e^{i\omega t}$, yields 
\begin{equation}
\left(\omega_{\theta}^{2}-\omega^{2}\right)\left(\omega_{z}^{2}-\omega^{2}\right)=\frac{\epsilon^{2}}{4mI}
\label{Eqmotion_fourthorder}
\end{equation}
The normal (eigen) frequencies of the coupled pendulum $\omega_{1}$ and $\omega_{2}$ are given by the two solutions of Eq.~\ref{Eqmotion_fourthorder}, yielding
\begin{equation}
    \begin{aligned}
    \omega_{1}^{2}= & \frac{\omega_{\theta}^{2}+\omega_{z}^{2}}{2}+\frac{1}{2}\left[\left(\omega_{\theta}^{2}-\omega_{z}^{2}\right)^{2}+\frac{\epsilon^{2}}{mI}\right]^{1/2}\\
    \omega_{2}^{2}= &\frac{\omega_{\theta}^{2}+\omega_{z}^{2}}{2}-\frac{1}{2}\left[\left(\omega_{\theta}^{2}-\omega_{z}^{2}\right)^{2}+\frac{\epsilon^{2}}{mI}\right]^{1/2}.
    \end{aligned}
    \label{eq:omegasFull}
\end{equation}
Furthermore, introducing the perfect tuning hypothesis $\omega_{\theta}=\omega_z=\omega$, i.e., assuming that the fundamental frequencies for both oscillation types are equal, the normal frequencies read
\begin{equation}
    \begin{aligned}
    \omega_{1}^{2}=&\omega^{2}+\omega\omega_{B}=\omega^{2}\left(1+\frac{\omega_{B}}{\omega}\right)\\
    \omega_{2}^{2}=&\omega^{2}-\omega\omega_{B}=\omega^{2}\left(1-\frac{\omega_{B}}{\omega}\right),
\end{aligned}
    \label{eq:omegasrewritten}
\end{equation}
where we have introduced the beat frequency
\begin{equation}
\omega_{B}^{2}=\frac{\epsilon^{2}}{4\omega^{2}mI}.
\label{eq:FreqStrongHyp}
\end{equation}
Note that this equation allows for the calculation of $\epsilon$ from experimental measurements of $\omega$, $\omega_B$, $m$ and $I$. Commonly, Eq.~\ref{eq:omegasrewritten} is simplified by a binomial expansion for $\nicefrac{\omega}{\omega_{B}}\ll1$: 
\begin{equation}
\begin{aligned}
\omega_{1}=\omega+\omega_{B}/2\\
\omega_{2}=\omega-\omega_{B}/2
\end{aligned}.\label{eq:beatingmodes}
\end{equation}
Hence, $\omega$ represents the average normal frequency, and the beat frequency $\omega_B$ is the difference between the normal frequencies.

\subsection{Optimum beating conditions }\label{CItuned}

Following \cite{berg1991}, we can condense the vertical $z\left(t\right)$ and angular $\theta\left(t\right)$ components of the movement in the vector $\hat{X}(t)=z\left(t\right) \hat{z}+\theta\left(t\right)\hat{\theta}$, where $\hat{z}$ and $\hat{\theta}$ are versors in the vertical and horizontal directions of motion, respectively. When the initial velocities are $\dot{z}=0$ and $\dot{\theta}=0$, the solution of the motion equations is given by the following superposition of normal modes $\hat{\eta}_{1}$ and $\hat{\eta}_{2}$
\begin{equation}
    \hat{X}(t)=A_{1} \hat{\eta}_1 \textrm{cos}\,\omega_{1}t+A_{2} \hat{\eta}_2\textrm{cos}\,\omega_{2}t.
    \label{eq:normalmodes}
\end{equation}  
The normal modes are given by
\begin{equation}
    \begin{aligned}
        \hat{\eta}_{1}= & \hat{\theta}+\hat{z} \sqrt{\frac{I}{m}}\\
        \hat{\eta}_{2}= & \hat{\theta}-\hat{z} \sqrt{\frac{I}{m}}.
    \end{aligned}
\end{equation}
The amplitude of the normal modes are given by the initial conditions $z(0)=z_0$ and $\theta(0)=\theta_0$, yielding
\begin{equation}
\begin{aligned}
A_{1}= &\frac{1}{2}\left[\theta_{0}+z_{0} \sqrt{\frac{m}{I}}\right]\\
A_{2}= &\frac{1}{2}\left[\theta_{0}-z_{0} \sqrt{\frac{m}{I}}\right]
\end{aligned}
\end{equation}
A quick inspection of the amplitudes reveals that the first mode is symmetric, i.e., rotation and translation are in phase, while the second mode is anti-symmetric. Beating takes place when both normal modes have the same amplitude, namely $A_{1}=\pm A_{2}$. One of the initial conditions that generates beating motion is  $\theta_{0}=0$, leading to the solutions
\begin{equation}
\begin{cases}
z\left(t\right)=z_{0}\;\textrm{sin}\,\omega t\;\textrm{sin}\,\frac{\omega_{B}}{2}t\\
\theta\left(t\right)=-z_{0}\sqrt{\frac{I}{m}}\,\textrm{cos}\,\omega t\;\textrm{cos}\,\frac{\omega_{B}}{2}t
\end{cases}.
\end{equation}
Under these conditions, the pendulum oscillates at the resonance frequency $\omega$, and the oscillation is modulated by the beating frequency $\omega_B$. Moreover, the rotational and translational oscillations have a  $\pi/2$ phase difference, while the total energy remains constant. 
In order to compare the solutions to the motion equations to experimental data, it is worth noting that smartphones measure translational and rotational accelerations through their 3D accelerometers and gyroscopes, respectively \cite{christoph2020lab}. For the motion described by Eq.~\eqref{eq:normalmodes}, the corresponding accelerations are
\begin{equation}
    \ddot{\hat{X}}(t)=-\omega_{1}^{2}A_1\hat{\eta}_{1}\textrm{cos}\;\omega_{1}t-\omega_{2}^{2}A_2\hat{\eta}_{2}\textrm{cos}\,\omega_{2}t.
    \label{eq:acceleration}
\end{equation}
Hence, the normal modes in acceleration conserve the same phase relationship than the displacements. However, it is important to note that the initial conditions for beating, $z_{0}=0$  or $\theta_{0}=0$ (with null initial velocities), do not yield accelerations with equal amplitudes for both frequencies, as it does with positions, but rather $A_{1}/A_{2}=\omega_{1}^{2}/\omega_{2}^{2}$. In other words, the initial conditions that are necessary to observe beatings in the positional coordinates do not lead to beatings in the accelerations. 

From Eq.~\eqref{eq:acceleration}, we observe that imposing a zero-initial-torque condition leads to beating in acceleration. Such condition be readily imposed experimentally by simply lifting the pendulum with one finger under its center of mass, up to the position in which the torque vanishes. The finger's position must enable free rotation. The initial angle for such condition can be found by imposing $\ddot{\theta}=0$ in Eq. \eqref{Eqmotion}, leading to ${\omega^{2}\theta_0=-\frac{\epsilon}{2I}z_0}$. Under these conditions, the observed accelerations are
\begin{equation}
\begin{cases}
\ddot{z}(t)=-\omega^{2}\left(1-\frac{\omega_{B}^{2}}{\omega^{2}}\right)z_{0}\,\textrm{cos}\,\omega t\;\textrm{cos}\,\frac{\omega_{B}}{2}t \\
\ddot{\theta}(t)=\omega^{2}\left(1-\frac{\omega_{B}^{2}}{\omega^{2}}\right)z_{0}\sqrt{\frac{m}{I}}\,\textrm{sin}\,\omega t\;\textrm{sin}\,\frac{\omega_{B}}{2}t
\end{cases},
\label{MotionFreeTorque}
\end{equation}
thus showing perfect beating. On the other hand, the observed positions for the same initial conditions are
\begin{equation}
\begin{cases}
z(t)=z_{0}\;\left(\textrm{cos}\,\omega t\;\textrm{cos}\,\frac{\omega_{B}}{2}t+\frac{\omega_B}{\omega}\textrm{sin}\,\omega t\;\textrm{sin}\,\frac{\omega_{B}}{2}t\right)\\
\theta(t)=-z_{0}\sqrt{\frac{m}{I}}\left(\textrm{sin}\,\omega t\;\textrm{sin}\,\frac{\omega_{B}}{2}t+\frac{\omega_B}{\omega}\textrm{cos}\,\omega t\;\textrm{cos}\,\frac{\omega_{B}}{2}t\right)
\end{cases}.
\end{equation}
We can see that, in addition to the main beating term, the positions have a term in anti-phase with respect to the main one, and proportional to $\frac{\omega_{B}}{\omega}$. As a consequence, the envelope of the positions has a minimum of $\frac{\omega_{B}}{\omega}$ instead of zero.

\subsection{Modelling attenuation}

The oscillations of a real Wilberforce pendulum are attenuated by friction. To model attenuation we can add damping terms as proposed in \cite{devaux2019cross}, to obtain the motion equations
\begin{equation}
\begin{cases}
m\ddot{z}+\alpha_z\dot{z}+kz+\frac{\epsilon}{2}\theta=0\\
I\ddot{\theta}+\alpha_\theta\dot{\theta}+\delta\theta+\frac{\epsilon}{2}z=0
\end{cases},
\label{Eqmotion2}
\end{equation}
Assuming a harmonic solution, Eq~.\eqref{Eqmotion2} is equivalent to Eq.~\eqref{Eqmotion} for imaginary elastic constants $\boldsymbol{k}=k+i\alpha_z$ and $\boldsymbol{\delta}=\delta+i\alpha_\theta$. For the sake of simplicity, we assume that the intensity of the rotational and translational damping are equal, namely    $\zeta=\frac{\alpha_z}{2m\omega}=\frac{\alpha_\theta}{2I\omega}$. Note that the natural frequency of both the translational and rotational oscillations is modified by a factor $\sqrt{1+i2\zeta}$, which can be approximated to $1+i\zeta$ when the attenuation is small enough, i.e., $\zeta<<\omega$. We thus expect the analysis of Sec. I.A. to remain valid even when attennuation is considered. In particular, Eqs.~\eqref{Eqmotion_fourthorder} to \eqref{eq:FreqStrongHyp} are still valid for a complex harmonic frequency $\boldsymbol{\omega}=\omega\left(1+i\zeta\right)$. At the first order, the normal frequencies have the same complex term $\boldsymbol{\omega}_{1,2}=\omega_{1,2}\left(1+i\zeta\right)$. In consequence, the time solutions of section \ref{CItuned} are multiplied by a  term $e^{-i \zeta t}$ to include attenuation.

\section{Experiment}\label{sec:expe}

The mobile part of the pendulum depicted in Fig.~\ref{fig:Sketch} consists of a metallic cylindrical box with a diameter \SI{9.5}{cm}. In addition, the box has two holes with two \SI{10.5}{cm} bolts and 4 screws to increase the moment of inertia. The box is attached to a \SI{7.0}{cm}-diameter spiral aluminium spring, with a $2\times$\SI{1}{mm^2} rectangular section. Only 2/3 of the total spring length was used (about 80 loops). The spring is attached to the ceiling. We tape a $16\times$\SI{8}{cm^2} smartphone to the box facing the ground, thus enabling access to the tactile screen.
All spatial dimensions were measured with a caliper and all masses with a kitchen scale.  The total momentum of inertia of the system ${I}$ was calculated adding the momentum of inertia of the different parts independently calculated assuming homogeneous density. For the smartphone, we assume its momentum of inertia to be that of a rectangular box. The bolts and screws are consider as thin rods, and the box as a cylinder shell. Hence, the total inertia is ${I=\left(1925\pm6\right) \SI{}{\kilogram / \metre \squared}}$, where the uncertainty corresponds to the propagation of the uncertainty in the mass and spatial dimensions. The total pendulum mass is ${m=\left(398.9\pm0.5\right)10^{-3}}\SI{}{\kilogram}$. 

For data acquisition, the smartphone records both accelerations and angular velocities in 3D as indicated in. Fig.~\ref{fig:Sketch}. We use the application $\text{Phyphox}^{\text{\textregistered}}$
to read the smartphone accelerometer and gyroscope with a sampling frequency of $f_s = \SI{0.1}{\hertz}$ \cite{christoph2020lab}. 
Phyphox allows to save the the data in .csv format and e-mails it instantaneously for further analysis on a PC, for which we build a very simple pipeline on Python.  

\subsection{Estimating the position of the accelerometer}

\begin{figure}[h]
\centering \includegraphics[width=1\columnwidth]{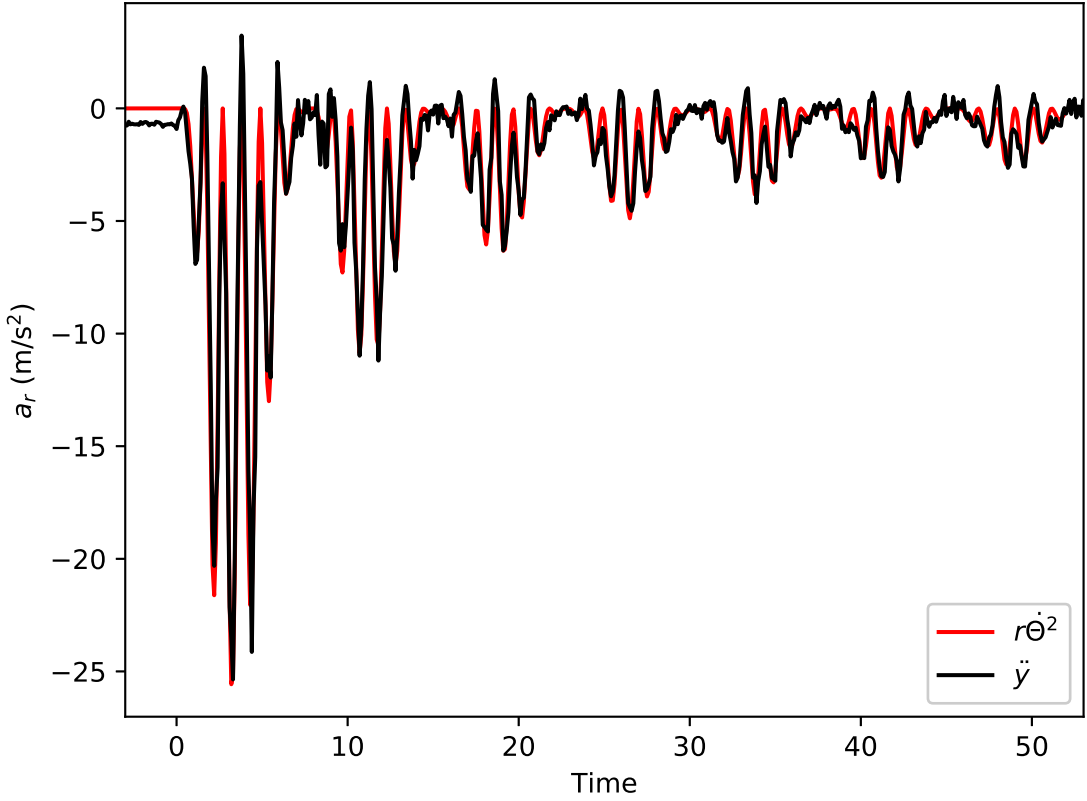}
\caption{Radial acceleration on the $xy$ plane measured from the accelerometer and gyroscope outputs. Black line: acceleration along the $\vec{y}$ direction. This acceleration corresponds to the radial acceleration in the cylindrical coordinate $\ddot{\vec{u_r}}$. Red line: $r\dot{\theta}^2 $  computed from the angular velocity $\dot{\theta}$ measured by the gyroscope.}
\label{cilinCoor}
\end{figure}

One of the advantages of our experimental setup over experiments based on video recordings, is having easy access to a frame of reference attached to the pendulum. This allows for simultaneous measurements of the acceleration along the perpendicular plane to the spring's main axis by two different methods, which can be used to estimate the position of the accelerometer. The acceleration in the horizontal plane can be measured either by the accelerometer, which expresses it in the Cartesian coordinates $x$ and $y$ depicted in Fig.~\ref{fig:Sketch}, or by the $r,\theta$ cylindrical coordinates, where $\theta$ is the output of the gyroscope. The relationship between the accelerations lead to the estimation of the coordinates $(r_{acc},\theta_{acc})$ of the accelerometer.  

We can write the acceleration along the smartphone's $y$ axis of Fig.~\ref{fig:Sketch} in terms  of the cylindrical coordinates: 

\begin{equation}
    \begin{aligned}
        \vec{y}= & \textrm{sin}\theta_{acc}\;\hat{u}_r+\textrm{cos}\theta_{acc}\;\hat{u}_\theta\\
        \ddot{\vec{y}}= &-r_{acc} \dot{\theta}^2 \textrm{sin}\theta\; \hat{u}_r+r_{acc} \ddot{\theta}\textrm{cos}\theta\;\hat{u}_\theta
    \end{aligned}
\end{equation}
, where $\hat{u}_r$ and $\hat{u}_\theta$ are the radial an angular cylindrical directions, respectively. When plotting the acceleration along the $y$ axis for the pendulum, we observe that it is always negative (see black line in Fig.~\ref{cilinCoor}), which shows that $\theta_{acc}\approx \pi/2$, i.e., the accelerometer is positioned along the $y$ axis. Hence, we can relate the measurements of the accelerometer and gyroscope by 
\begin{equation}
    \ddot{\vec{y}}=-r_{acc}\dot{\theta}^2 \hat{u}_r.
    \label{radialAcc}
\end{equation}
The value of $r_{acc}$ can be found then by minimising the distance between the signals for $\ddot{y}$ and $r_{acc}\dot{\theta}^2$ for the motion of the pendulum \cite{larnder2020purely}. Figure \ref{cilinCoor} shows in red the signal of $r_{acc}\dot{\theta}^2$ that minimises such distance. We estimated through this method $r_{acc}\approx \SI{3.3}{\cm}$. Note that this process is independent of the initial conditions set for the pendulum, however it is highly sensitive to the orientation of the smartphone with respect to the rotational axis. Thus, careful alignment is recommended.  

\subsection{Normal modes}
\begin{figure}[h]
\centering \includegraphics[width=1\columnwidth]{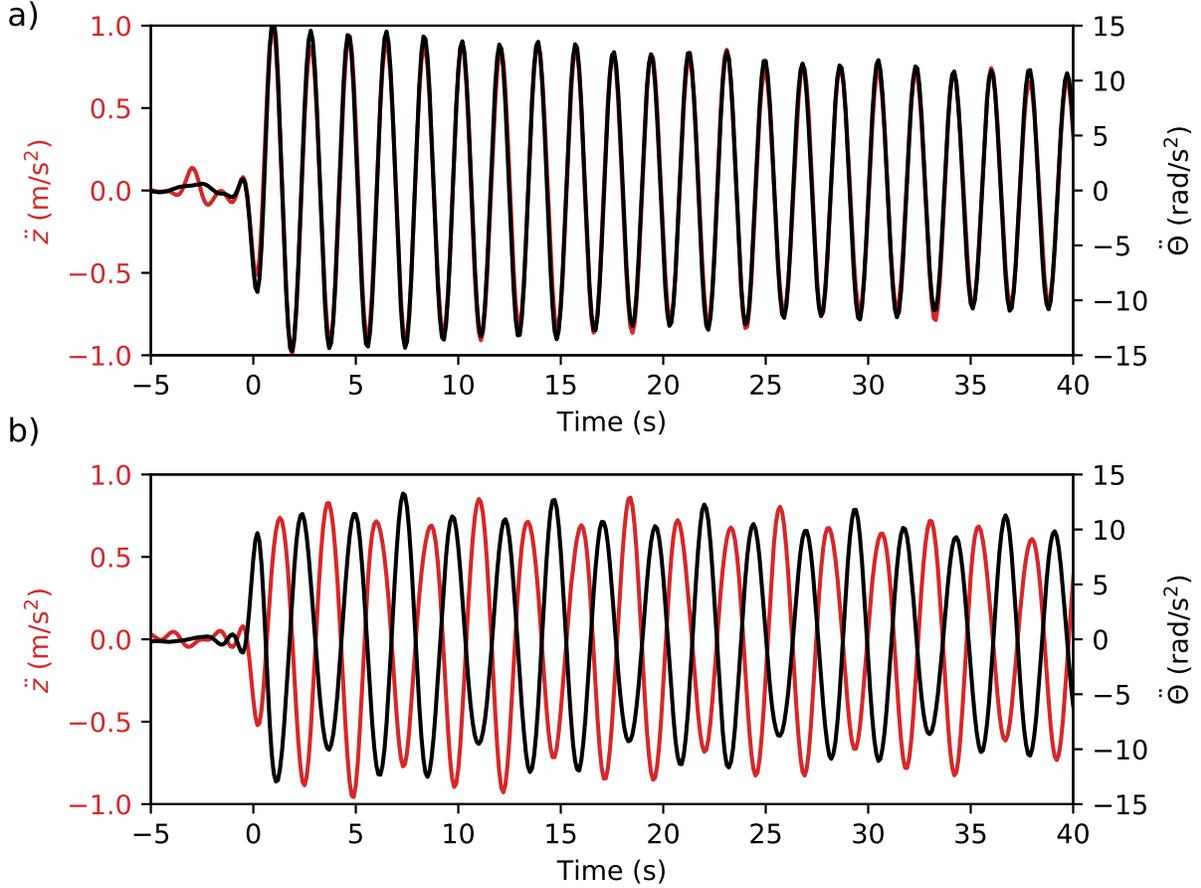}
\caption{\label{fig:Modes} Normal modes of the Wilberforce pendulum observed experimentally in the acceleration signals. a) Symmetric (first) mode observed in the accelerations $\ddot{z}(t)$ (red line) and  $\ddot{\theta}(t)$ (black line). b) Anti-symmetric (second) mode.}
\end{figure}
As shown by Berg and Marshal \cite{berg1991}, the conditions ${z_0=\pm\theta_0\sqrt{\nicefrac{I}{m}}}$ set the system to oscillate in the symmetric and asymmetric normal modes (see Sec.~\ref{sec:theory}. B), with $z_0=0.1 \SI{}{\meter}$ . For such initial conditions, the simultaneous rotational and translational accelerations are shown inf Fig.~\ref{fig:Modes}, where we can see them in phase for the symmetric mode (panel a)) or in anti-phase for the asymmetric mode (panel b)).  Here, we calculate the rotational acceleration as the numerical gradient of the angular velocity measured by the smartphone. For both acceleration signals, we remove the DC components and apply a band-pass frequency filter between 0.1 to 1 Hz\footnote{The scipy.signal has a \textit{detrend} function, a Butterworth filter coefficient design (\textit{butter}) and  [textit{sosfiltfilt} to apply the filter in the time domain}. Attenuation is clearly visible along the twenty cycles represented in Fig.~\ref{fig:Modes}.
We estimate the two normal modes frequencies from the discrete Fourier transform of the recorded signals for ten realisations with different initial conditions, obtaining $f_1=\frac{\omega_1}{2\pi}=0.539 \pm 0.003 \SI{}{\hertz}$ and $f_2=0.407\pm0.001 \SI{}{\hertz}$, respectively. The uncertainties correspond to their standard deviations among realisations. The fundamental and beating frequencies can be estimated from $f_1$ and $f_2$ via Eq. \eqref{eq:inverspuls}.

\subsection{Beating observation}

\begin{figure}[h]
\centering \includegraphics[width=1\columnwidth]{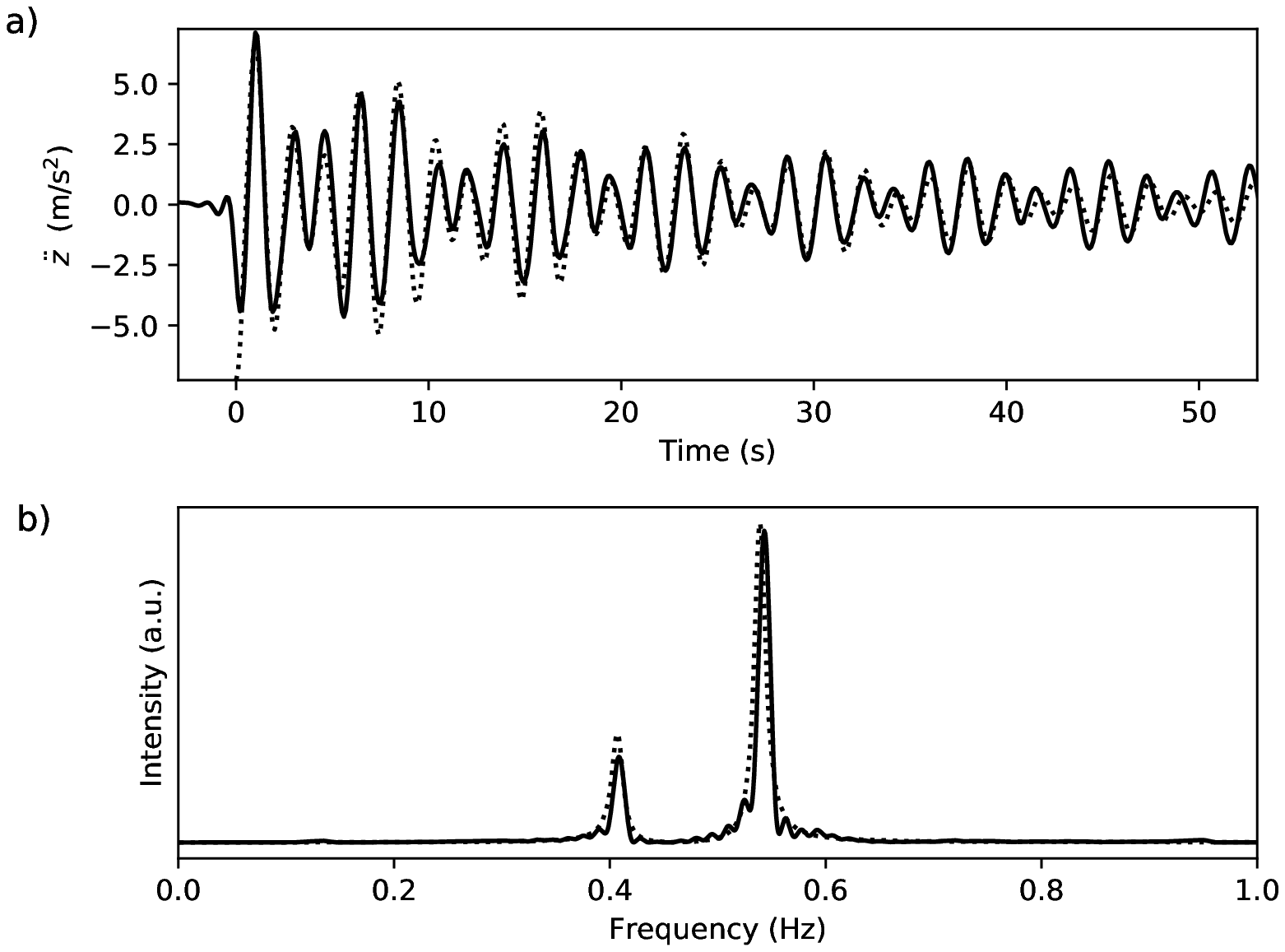}
\caption{ Acceleration as a function of time for initial conditions that ensure beating in the positional signals. a) Acceleration along  the $\vec{z}$-axis for initial conditions $z_0=\SI{0.82}{\metre}$ and $\theta_0=0$. The acceleration measured experimentally (black line) is in full line with numerical experiments (dotted line). b) Corresponding spectral densities. The ratio between the amplitudes of the first and second resonances are $\omega_{1}^{2}/\omega_{2}^{2}$, as per Eq.~\eqref{eq:acceleration}.
}
\label{fig:CI1}
\end{figure}

As we showed analytically in Sec.~\ref{subsec:motionequations}, setting the initial conditions that correspond to beating for the translational and rotational positions, namely $z_0= \SI{0.82}{\metre}$ and $\theta_0=0$, lead to partial beatings in the accelerations $\ddot{z}(t)$ and $\ddot{\theta}(t)$ (black line in Fig.~\ref{fig:CI1} a)). Numerical simulations of the solution of the motion equations are described in  Sec.~\ref{subsec:numericalexperiments}. A good agreement with the experimental data can be observed for the accelerations as a function of time and spectral density (dashed lines in Fig.~\ref{fig:CI1} a) and b)). To calculate the spectral density, we compute the discrete Fourier transform after applying a Blackman window to minimize sidelobes. The amplitude ratio between the first and second modes of the simulation is given by $A_{1}/A_{2}=\omega_{1}^{2}/\omega_{2}^{2}=0.57$, as predicted in Sec.~\ref{CItuned}. We measure experimentally $A_{1}/A_{2}=0.52$. The peak frequency differences between experimental data and simulations are less than $0.5\%$. To obtain perfect beating on the acceleration time series, however, we need to impose the no-initial-torque condition (see Sec.~\ref{CItuned})

\begin{figure}[h]
\centering \includegraphics[width=1\columnwidth]{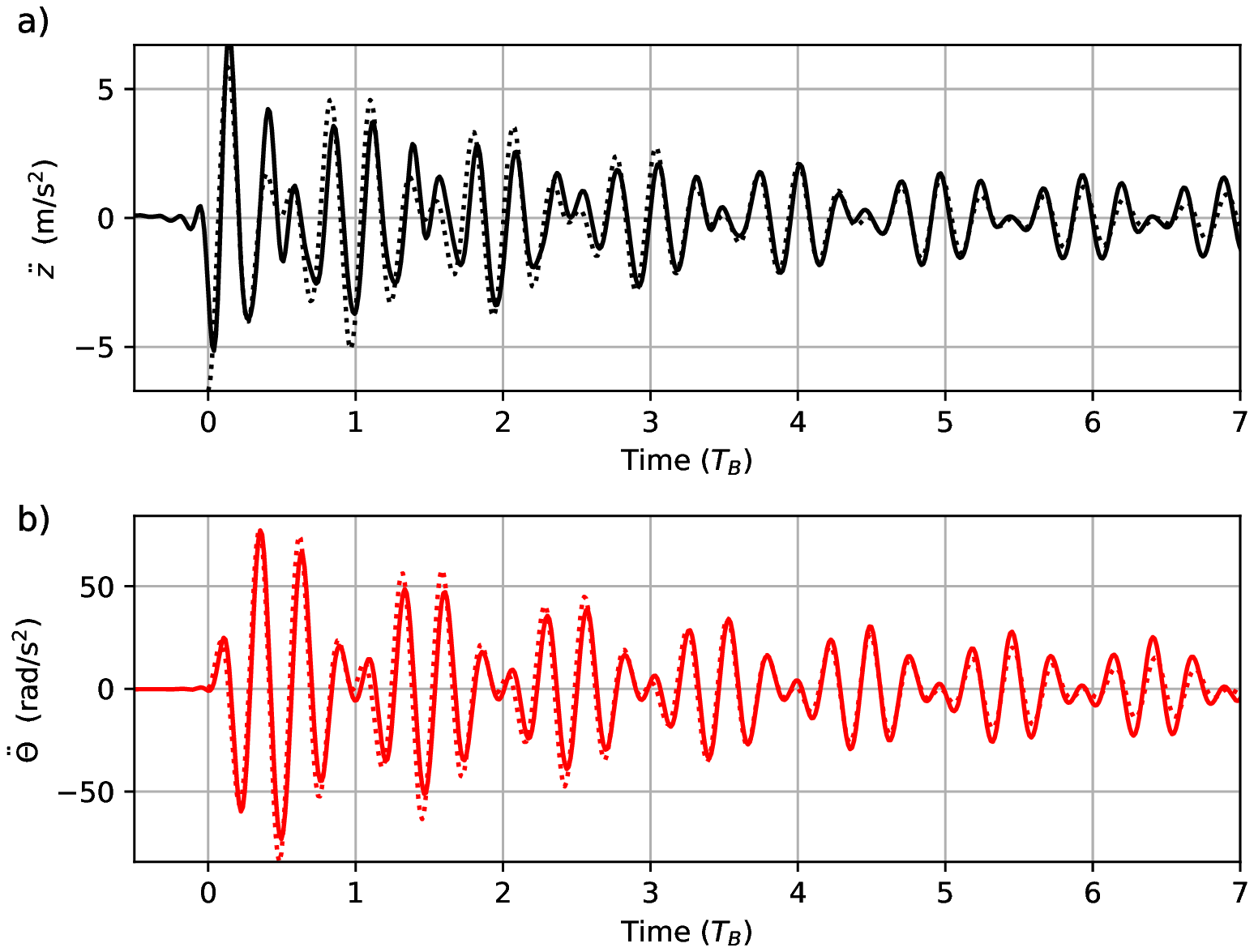}
\caption{ Translational and rotational accelerations for initial conditions that ensure beatings in the acceleration signals. a) Translational acceleration for initial conditions $z_0=\SI{0.82}{m}$ and no initial torque (${\theta_0=\SI{-3.2}{rad}}$). b) Rotational acceleration around $\vec{z}$-axis for the same initial conditionas as a). The time is rescaled to beating time $T_B=(7.6 t) \;\SI{}{\s}$, to facilitate the visualisation of the phase difference between the rotational and translational accelerations. The experimental measurments (full lines) are in great agreement with the numerical experiments (dotted lines).}
\label{fig:CI2_time}
\end{figure}
\begin{figure}[h]
\centering \includegraphics[width=1\columnwidth]{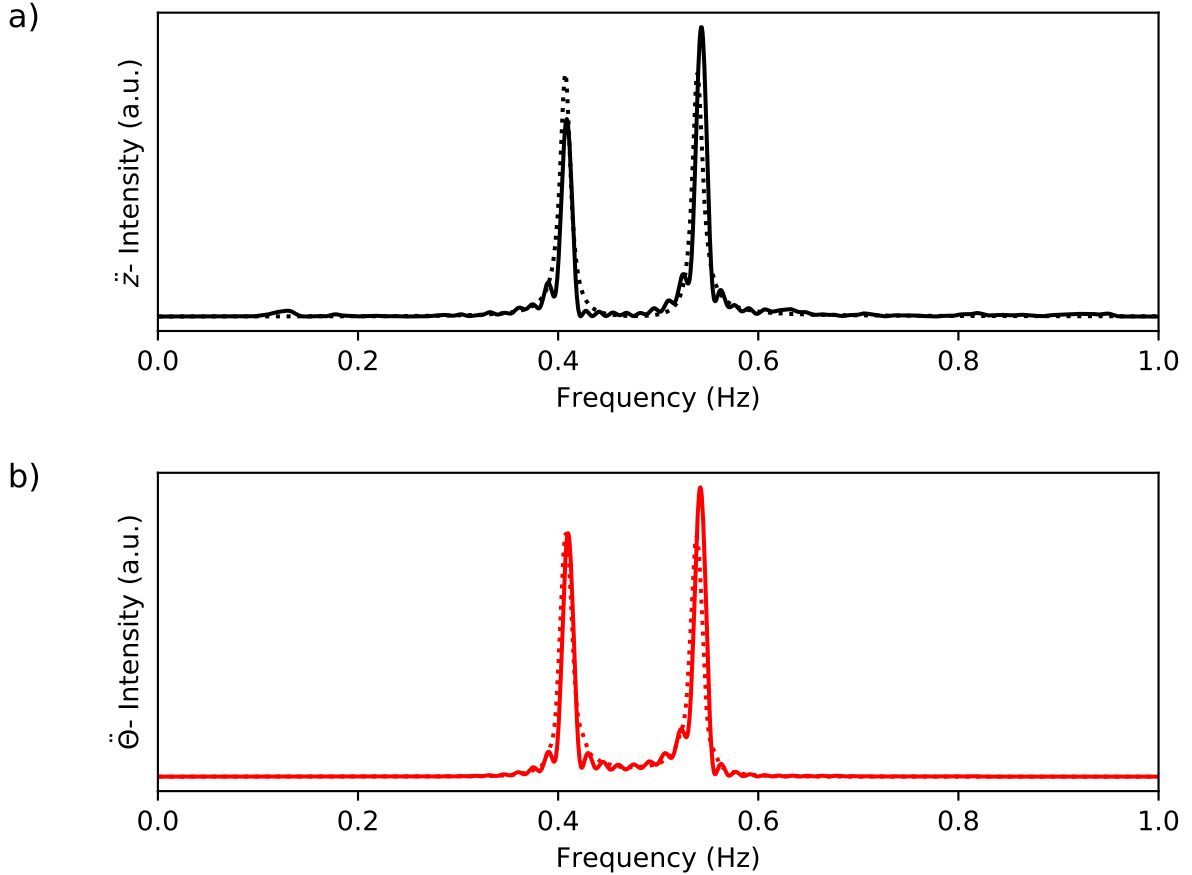}
\caption{\label{fig:CI2_freq} Spectral densities for initial conditions that ensure beatings in the translational (panel a)) and rotational (panel b)) accelerations. In the simulation (dotted lines), the amplitude of the first and the second resonances coincide, in agreement with Eq.~\eqref{MotionFreeTorque}. The experimental components (full lines) are slightly different, having a ratio of $0.82$.}
\end{figure}

When we set the initial conditions with no initial torque (in our case  $z_0=\SI{0.82}{\metre}$, $\theta_0=-\frac{\epsilon z_0}{2I\omega^{2}}=\SI{-3.2}{rad}$), we observe experimentally perfect beating in the rotational and translational accelerations, as shown in Fig.~\ref{fig:CI2_time}. Note that setting the initial angle $\theta_0$ to its theoretical value is not needed, but we rather set up the pendulum by lifting it with one finger until it can rotate freely, thus satisfying the no-initial-torque condition. Simulations of the pendulum's motion equations with the same initial conditions are in good agreement with the experimental data (dotted lines in Fig.~\ref{fig:CI2_time} a) and b)). The observed phase difference between the translational and rotational accelerations, which are in agreement with Eq.~\eqref{MotionFreeTorque}, illustrates the energy exchange between translational and rotational motions. Our simulations predict both normal frequencies to have the same contribution to the spectrum of the rotational and translational accelerations (dotted lines in Fig.~\ref{fig:CI2_freq} ). However, the experimental amplitude ratio between the first and the second mode is $A_{1}/A_{2}=0.82$, for both acceleration types. This disagreement between experiment and simulations can be explained either by a small difference between $\omega_{\theta}$ and $\omega_z$, or a small error on the initial angle due to friction.

\subsection{Numerical experiments \label{subsec:numericalexperiments}}

To solve numerically the set of Equations \eqref{Eqmotion2}, we re-write the autonomous system in the form $\dot{\bf{X}}=f(\bf{X})$, where $\bf{X}$ is a vector quantity. Since the coupling term in our system is linear with respect to the translational and rotational positions, the motion equations conform a linear system. Defining the new variables $r=\dot{z}$ and $s=\dot{\theta}$, the motion equations read $\dot{r}=-(\nicefrac{k}{m})z -(\nicefrac{\epsilon}{2m})\theta-(\alpha_z/m)r$, and $\dot{s}=-(\nicefrac{\delta}{I})\theta -(\nicefrac{\epsilon}{2I})z-(\alpha_{\theta}/I)s$. We can thus define $\bf{X}=(z,\theta,r,s)^{\textrm{t}}$, to express the motion equations as $\dot{\bf{X}}= \mathcal{M}\bf{X}$, where the matrix $\mathcal{M}$ is defined by
\begin{equation}
    \mathcal{M}=\begin{bmatrix}
    0 & 0 & 1 \quad& \quad0 & \\
    0 & 0 & 0 &\quad 1 &\\
    -\nicefrac{k}{m} & -\nicefrac{\epsilon}{2m} & -\alpha_z/m & \quad 0 & \\
    -\nicefrac{\epsilon}{2I} & -\nicefrac{\delta}{I} & 0 &\quad -\alpha_{\theta}/I &
    \end{bmatrix}.
\end{equation}
Naturally, the solution for this system is given by $\bf{X}(t)=e^{\mathcal{M}}\bf{X(0)}$, and an analytical solution can be expressed in terms of the eigenvalues and eigenvectors of $\mathcal{M}$. If the eigenvalues and eigenvectors of $\mathcal{M}$ are not simple to calculate, they can be computed numerically. 

An alternative way of solving the linear system is numerically integrating it using a Runge-Kutta method. Here, we choose to use this approach for two reasons: the Wilberforce pendulum is a great example to introduce numerical methods for ODEs to students, and this approach can still be used if we model the coupling non-linearly. When non-linear couplings are considered, approximate analytical solutions can still be obtained, for example using perturbation theory; however, this processes can be cumbersome and not well suited for introductory courses.

We apply a classical, order 4 Runge-Kutta method to numerically calculate the solution of $\dot{\bf{X}}= \mathcal{M}\bf{X}$. The Python code, available in GitHub, uses the package \textit{odeint} of \textit{scipy} 
 and is adapted from \cite{WebRefPython}. Note that the motion equations assume the same conditions of the theoretical analyses in Sec.~\ref{sec:theory}, including equal attenuation for the rotational and translational oscillations. We estimate the simulation parameters from the experimentally measured normal frequencies, mass and inertia, namely 
\begin{equation}
    \begin{aligned}
    \omega=&\sqrt{\frac{\omega_1^2+\omega_2^2}{2}},\\
    \omega_{B}=&\sqrt{\frac{\omega_1^2-\omega_2^2}{\omega}},\\
    \epsilon=&2\omega\omega_B\sqrt{mI}.
    \end{aligned}
    \label{eq:inverspuls}
\end{equation}
These expressions assume the simplification of Eq.~\eqref{eq:FreqStrongHyp} with $\omega_B/\omega=0.27$. Such simplification induces an error that is smaller than $1\%$. Nevertheless, this error accumulates period after period, and the time shift is visually noticeable after 10 periods. We thus strongly recommend to use, instead, the full expression of Eq.~\eqref{eq:inverspuls}. We estimate the damping factor  $\zeta=0.013$ is by minimising the difference between the numerical and experimental data.

\section{Summary and Discussion} 

In this work we have proposed a simple and low-cost setup for the study of the Wilberforce pendulum. By attaching a smartphone to the pendulum we measured the acceleration and angular velocity via the accelerometer and gyroscope of the device. Supported by numerical solutions to the motion equations (Sec.~\ref{sec:expe} D), we studied quantitatively the normal modes (Sec.~\ref{sec:expe} B) and beat conditions (Sec.~\ref{sec:expe} C) of the system obtaining a remarkable agreement between experiments and numerical results (Figures 4 to 7).

Notably, our experimental setup measures the motion with respect to a frame of reference attached to the pendulum. Such feature simplifies the analysis and sheds light into the optimum technique for setting the initial conditions to obtain perfect beatings in acceleration (Sec.~\ref{sec:theory} C) which, to the best of our knowledge, has not been shown before. In contrast, with setups based on filming the motion of the pendulum, measurements in the frame of reference of the pendulum can only be indirectly calculated from data obtained in the earth's frame of reference \cite{greczylo2002using}. 

The Wilberforce pendulum can be a valuable asset for laboratory courses or as a demonstration of coupled oscillations in university physics courses. The analysis of its motion involves critically reviewing key concepts in physical sciences, such as normal modes, coupled systems, cylindrical coordinates or relative motion. Moreover, the experimental study of the pendulum can serve as an ideal gateway to learn data analysis techniques such as the fast Fourier transform or data filtering, as well as to numerical methods for ordinary differential equations. 
      
The simplicity of our experimental setup stems from the use of a smartphone as our measuring device, rendering the study of the Wilberforce pendulum easily reproducible. All the numerical tools we use in this study are freely available online. In addition, we provide data sets for further analysis, alongside with the Python scripts we used to analyse the experimental data and perform numerical experiments.

\subsection*{Data accessibility} 
Numerical simulation, data sets and data analysis are freely available at \href{https://github.com/RodrigoGarciaTejera/WilberforcePendulum }{GitHub repository}.

\subsection*{Conflict of interest declaration} 
The authors have no conflicts to disclose.


%

\bibliographystyle{unsrt}
\bibliography{biblio}

\end{document}